# LOW EARTH ORBIT SATELLITES PROVIDE CONTINUOUS ENTERPRISE DATA CONNECTIVITY

## A PREPRINT

**Sean P. Batir** *
IT Research & Innovation
BMW Group

**Nicholas R. Humann**
OneWeb
London, UK

**Carmel Ortiz**
OneWeb
London, UK

**David Bettinger**
OneWeb
London, UK

**Susanne Heger**
IT Research & Innovation
BMW Group

**Bennie Vorster**
IT Research & Innovation
BMW Group

July 24, 2020

## ABSTRACT

A critical problem in global telecommunication is the drastic increase in data volume transmitted and received across the world. To address the need for scalable telecommunication solutions in light of growing data volume, the BMW Group and Low Earth Orbit (LEO) satellite provider OneWeb pursue a proof of concept demo that assesses the potential of LEO networks for enterprise connectivity. Our results suggest that LEO satellite networks can enable the hybrid connectivity needed for continuous data transmission without interruption or loss of signal to enable the future of work and premium mobility. This is a proof of concept experiment to show throughput, latency, and key applications including handover to 4G and the use of a VPN while running cloud applications. Across three tests (i.e. entertainment and business productivity streaming services), the researchers demonstrate a 2-3x faster ping rate (ms), 4-5x faster download rates (Mb/s), and 30-60x (Mb/s) faster upload rates.

*Keywords* hybrid connectivity · LEO satellite · infotainment · backhaul · network infrastructure · future of work

## 1 Introduction and Background

In this research paper, we documented the development of a framework that could allow persistent, enterprise connectivity, even in the face of terrestrial network failure.

Roughly 20 years ago, the potential opportunities and limitations of low earth orbit (LEO) satellite technologies were demonstrated and identified. As an effective conduit for satellite communications, LEO constellations are both lower power and higher speed. LEO transmission has an average signal time delay on the order of only 5 to 10 msec. Likewise, the cost of establishing a link involves RF power budgets as low as 100 mW and a predicted, average power around 0.5W. [1] Additionally, the manufacturing of LEO satellites may provide a more scalable and economic solution for satellite based connectivity because less powerful, smaller launchers would be required to achieve lower orbit than traditional geostationary satellites.

However, LEO satellite technology also possesses barriers to entry. For instance, satellites in LEO orbit are continually in precession, acting as a moving target for ground stations. This demands highly directional antennas that can continually localize and beam-pass as different satellites move into visibility. Depending on altitude, some satellites may only be visible for 5 to 20 minutes per orbit, with an orbital period of roughly 90 minutes. Due to the ephemeral nature of LEO satellites disappearing along the horizon, LEO satellite systems require handling off-subscriber connections between satellites and their corresponding ground station, in order to provide continuous communication.

---

*Primary author correspondence: sbatir@alum.mit.edu

The window of altitude that LEO satellites are expected to orbit could provide data backhaul and continual connectivity, even in the event of a terrestrial network failure. The acceptable altitude range of (LEO) satellites is considered to exist between a lower bound of roughly 200-300 km, and an upper bound of approximately 2000km. At altitudes lower than 200 – 300km, the satellite may experience rapid orbital decay, while at altitudes higher than 2000 km, the integrity of the satellite may be compromised due to the intensity of trapped radiation belts. [2]

The maturing LEO satellite ecosystem offers a tangible solution to augment the increasingly congested space of terrestrial telecommunications. The demand for data consumption and transmission is rising, and some projections suggest the growth may follow an exponential trajectory during the 2020s. Likewise, a recent study regarding the expansion of 5G mobile telecommunication infrastructure suggested that while 5G small cell deployments are highly likely in urban and suburban locations, more cost-efficient wide-area coverage solutions are required to meet lower population density areas. [3]

This paper addresses the question, "Is it feasible to deliver reliable and high performance digital services through a hybrid satellite-terrestrial network?" The authors of this paper established a collaborative project between OneWeb, a low earth orbit satellite network provider; and BMW, a multinational company that produces vehicles and motorcycles. Our following experiments demonstrated successful audiovisual data streaming and teleconference transmission between nodes within both the internal BMW infrastructure and externally on the broader Internet. This was accomplished through OneWeb's operational, low earth orbit (LEO) network.

## 2 Materials and Methods

The experiments were carried out using 6 fully operational, OneWeb satellites that had reached a mission orbit of 1200km. These experiments lasted for approximately 15-20 minute testing windows, twice per day, every 12 hours in Melbourne, Florida. The testing was carried out over two weeks and in total, 8 tests were achieved. We designed our experiment to explore a hybrid system that passes data between WiFi provided by a LEO constellation, and a 4G/LTE terrestrial network due to the realistic utilization and near ubiquitous prevalence of urban and suburban 4G/LTE networks in 2020.

### 2.1 Low Earth Orbit Satellite System Components

The Low Earth Orbit satellite system produced by OneWeb consists of three key components: A Ka-Band terminal located 50 miles away from the test site in Clewiston, Florida; the first 6 operationally active, low earth orbit satellites; and corresponding Ku-band terminals, or functional ground antennas, manufactured by Intellian, located on the roof of the test facility located in Melbourne, Florida.

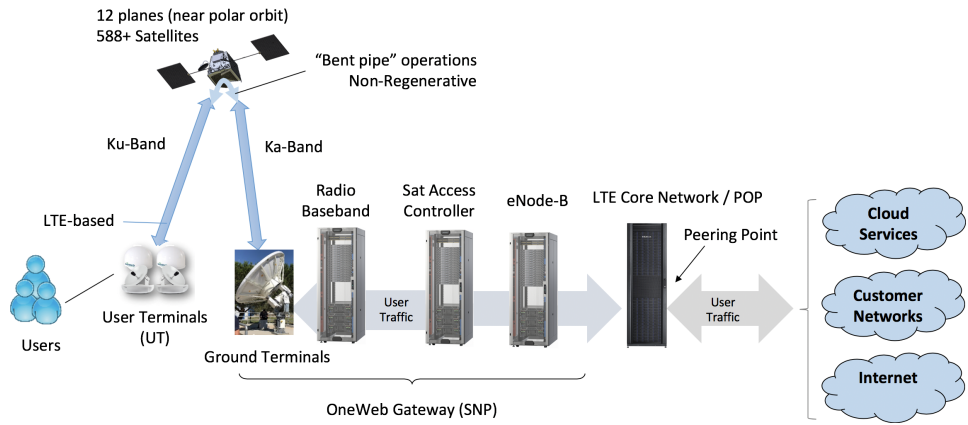

Figure 1: OneWeb bent pipe generation 1 system architecture.

Figure 1 illustrates the OneWeb bent pipe system architecture. In the experiment referenced in this study, preliminary results are derived from a network that had 6 fully operational LEO satellites, as of March 3, 2020.

### 2.1.1 Enterprise Connectivity Infrastructure

Test devices used in an enterprise connectivity experiment include a standard, corporate-issued Macbook Pro's 15" Late 2017 Model, two dedicated Satixfy modems to handle byte transfer (Rx + Tx) between the Ku band antennas, and a local WiFi router (Linksys E2500 N600 Dual-Band WiFi Router). Cellular switching tests were performed with a T9 T-Mobile 4G/LTE Hotspot, placed 12" inches from the test device.

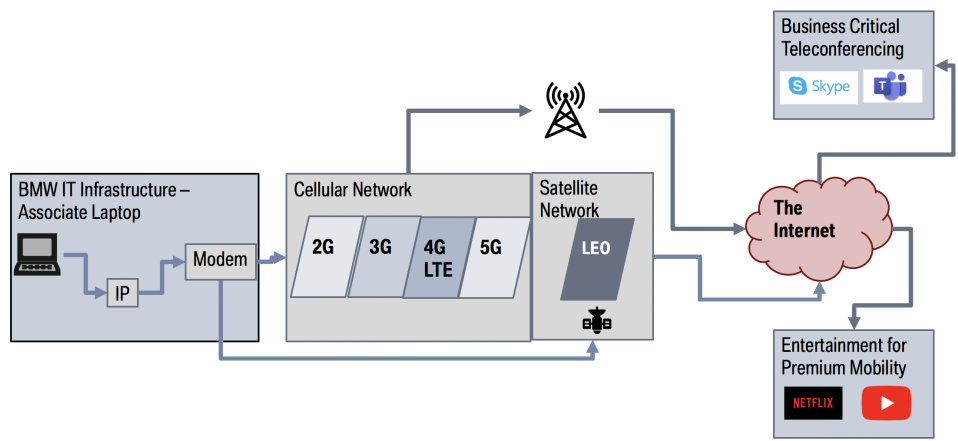

Figure 2: The overall experiment design that tests the connectivity between a component of the corporate infrastructure with Internet provided services, through hybrid connectivity between standard cellular network solutions and the LEO satellite network

Figure 2 captures the overall experiment design, which begins with a component of the BMW IT infrastructure, an associate laptop and demonstrates hand-off between a cellular network and a satellite network, in order to access Internet services across business critical teleconferencing, the AWS accessible cloud, and entertainment for premium mobility.

## 2.2 Standardizing Network Performance

The primary method used to establish standardized performance measurements was the execution of iPerf, a network performance measurement and tuning program written in C and executed in the command line. User Datagram protocol (UDP) data streams reveal results of datagram size, throughput and packet loss. For the remaining 6 tests, we sought to standardize network performance evaluation by using consumer-initiated testing by Ookla's command line interface tool, Speedtest. The speedtest command was executed within 1 second of starting the three streaming services (Netflix, YouTube, and Microsoft Teams). The command was executed after switching networks from LEO to 4G/LTE and from 4G/LTE to LEO.

Figure 3 captures the test interface used to assess network traffic in a realistic work and infotainment environment.

## 3 Results

The results reported below compare the ping rate (ms), download rate (Mb/s), and upload rate (Mb/s) across both the LEO and 4G/LTE communication protocols.

In addition to infotainment use cases (Table 1, and Table 2), we also confirm if business productivity tools such as Microsoft Teams and Skype, could provide continual connectivity.

Below, in Figures 4 and 5 we demonstrate successful transmission of business critical work applications, including Microsoft Teams and Skype, two communication tools used by the BMW Group NA.

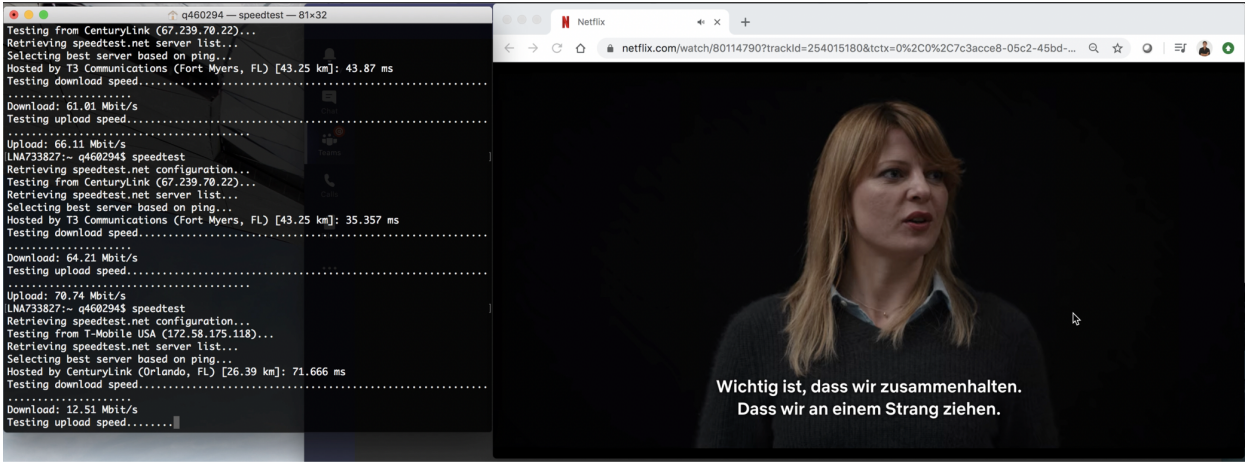

Figure 3: The left illustrates the use of the Speedtest package in Terminal to assess network traffic parameters shared in Table 1. The right illustrates a clip of seamlessly streaming the German Netflix show, Dark through the OneWeb LEO network.

Table 1: Netflix Streaming, Dark (2020)

| Performed Tests | Ping Rate (ms) | Download Rate (Mb/s) | Upload Rate (Mb/s) |
|---|---|---|---|
| LEO | 35.36 | 64.21 | 70.74 |
| 4G/LTE | 71.66 | 12.51 | 1.97 |

Table 2: YouTube Streaming, BMW i4 (2020)

| Performed Tests | Ping Rate (ms) | Download Rate (Mb/s) | Upload Rate (Mb/s) |
|---|---|---|---|
| LEO | 43.87 | 61.01 | 66.11 |
| 4G/LTE | 91.40 | 12.27 | 1.69 |

Table 3: Microsoft Teams (External)

| Table 3: | | | |
|---|---|---|---|
| Performed Tests | Ping Rate (ms) | Download Rate (Mb/s) | Upload Rate (Mb/s) |
| LEO | 37.85 | 80.07 | 64.76 |
| 4G/LTE | 91.40 | 12.27 | 1.69 |

## 4 Discussion

The results of the proof-of-concept demo suggest that LEO networks may provide a plausible backhaul solution, or network augmentation as streaming and remote work become more prevalent in the context of Force Majeurs such as the recent coronavirus pandemic of the early 2020s. The quantitative reduction in ping time and increased download and upload rates offer a qualitatively observable improvement in streaming service and user-experience. The authors acknowledge that some of the results may stem from the lack of network congestion, due to the sparse and underutilized nature of the 6 satellite operational LEO system. The OneWeb LEO prototype network, services rendered over the BMW infrastructure could maintain continuity and network stability through WiFi access to the LEO network.

The authors chose to execute Ookla's speedtest command line interface to realistically measure the user's streaming experience during use. In addition to the observed improvement in download and upload rates, the researchers also observed that the LEO network was able to provide a sufficiently long buffer-window of streamed data over Netflix and YouTube in order to successfully compensate for any lost data during momentary switching between 4G/LTE cellular connectivity and LEO connectivity. In future implementations, a possible improvement may be the automation of network switching based on real-time analysis of packet loss and online network traffic reporting.

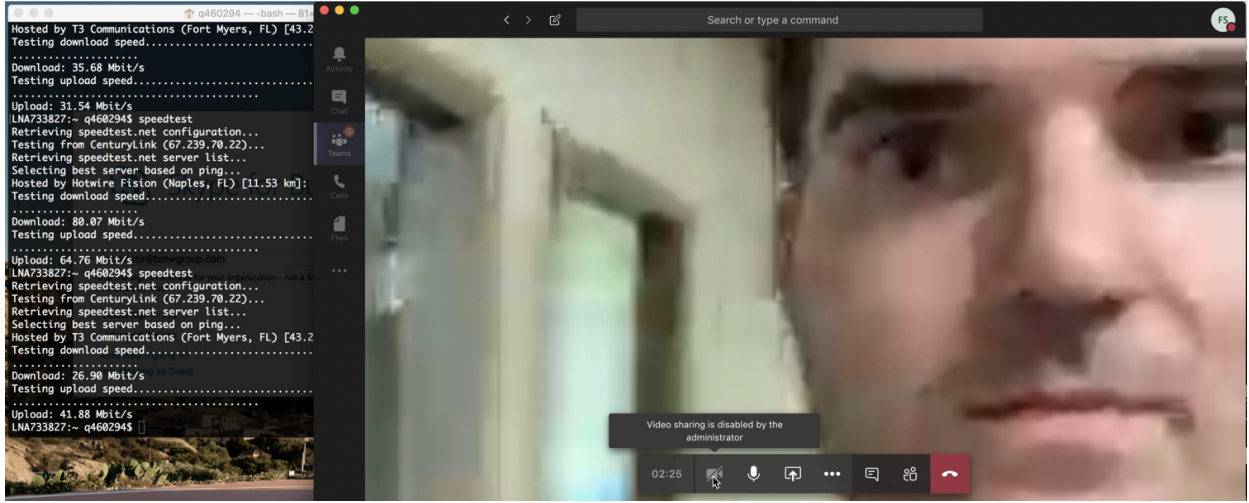

Figure 4: A screen capture that illustrates the use of the OneWeb LEO network for corporate video conferencing with external clients. The sender is from within the BMW network, and the receiver is from OneWeb.

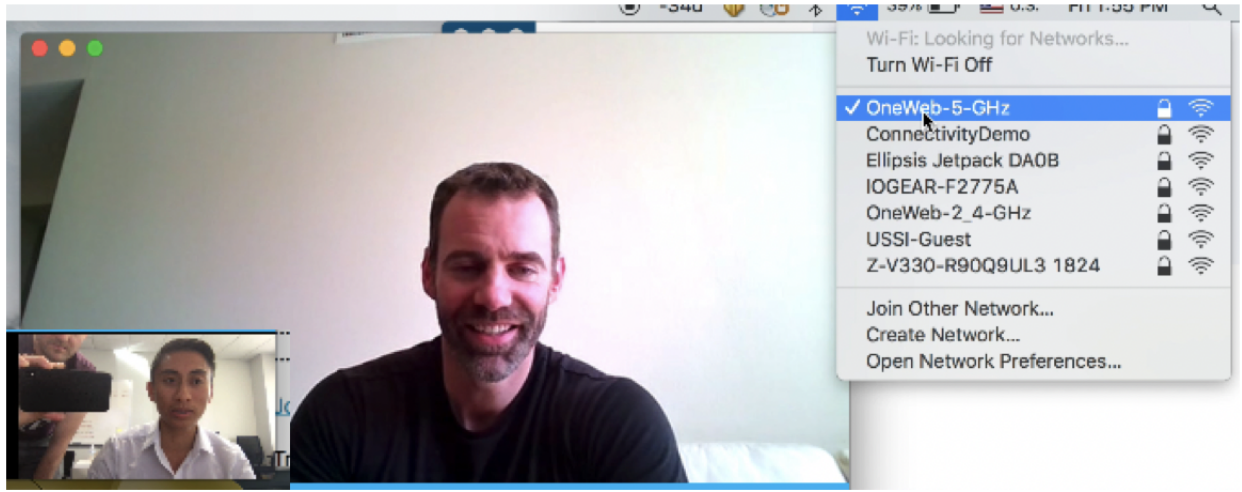

Figure 5: A screen capture that illustrates the use of the LEO network for internal BMW conferencing. While security measures within the BMW VPN network prevented server ping for upload and download rates, the high quality of the audiovisual feed is observed.

The authors emphasize that results were also constrained due to hardware limitations in this proof-of-concept experiment. Speeds would normally be higher on an integrated antenna dedicated for the purpose of enterprise or commercial transmission. To compensate for edge-case link conditions such as level changes and varying satellite altitudes during the 15 minute experiment window, data download rates were fixed at 60 Mb/s as a starting point. This was accomplished by operating the terminals in constant coding and modulation (CCM) mode, with low modulation and coding pairs (MODCOD).[4]

In the future, the use of LEO dedicated antennas could increase data throughput. Additional limiting factors include delivery of streaming services through WiFi that could have been throttled by a low-end WiFi router. Thus, while network congestion could make the speed slower, the technical set-up of the demo also could greatly alter the observed speed from the true capabilities of LEO networks.

While the experiment was limited by the available time window(2 slots of 15-20 minutes every 24 hour period), such a limitation fades away as more LEO satellites are launched, providing more coverage and accessible look-angles for corresponding ground stations to establish a connection and subsequent connectivity. At the time of this experiment, OneWeb had 6 fully operational satellites. Suggested future research directions include testing the compatibility of a LEO network with emerging 5G cellular networks. Likewise, the authors acknowledge that our approach was limited to connectivity in a generic enterprise IT system. Thus, a promising pathway of further research may involve the testing of continuous data transmission and streaming through a hybrid LEO-terrestrial cellular network within a test vehicle, instead of an isolated, enterprise laptop.

To provide measurable safety, reliable infotainment and productivity solutions, and potentially autonomous vehicle infrastructure, connected vehicles will require seamless handover between terrestrial and satellite connectivity. This will enable hardware, software, and experiences provision. Some potential vehicular use cases include software updates, connection to the cloud, maintaining connectivity in unconnected areas, emergency contact, weather reporting, navigation, driver assistance and monitoring, advanced safety, adherence to laws, border controls and regulations, vehicle monitoring, remote inspections, camera uploads, in vehicle entertainment, load monitoring, and emergency repairs.

## 5 Long Term Impact

The recent COVID-19 pandemic has highlighted the continually growing need for stable, global IT infrastructure. Across the United States, large sections of suburban areas do not yet have broadband due to the high cost of cable networks in the last 10 mile reach. Low Earth Orbit satellites provide an opportunity for access to low cost talent of coders and teleworkers, enabling new distributed working models and enhancing access to career opportunity across the United States. In the longer term, the use of LEO networks to support IT infrastructure and remote work can create new jobs and subsequently uplift educational opportunities in previously untouched territories. The subsequent increase in education and anticipated reduction of crime will enable LEO networks to reduce inequity both within the United States and across the world.

## 6 Conclusion

With the turn of a new decade, the ability to remain connected despite physical location and barriers will only become more paramount. A potential consequence of increased autonomous vehicles on the roads will be the sharp increase in demand for continuous connectivity. While some autonomous vehicle functionality will be relegated to edge-device and vehicular processing capabilities, governments and consumers may require a high standard and quality of continuous connectivity for the vehicle to understand holistic contextual information and provide safe navigation. The results in this paper illustrate that a hybrid satellite-terrestrial network is not only promising, but also possible and provides high performance data transmission. This vision may be readily achieved if a complete LEO satellite constellation and corresponding network of ground terminals can emerge that is capable of maintaining LEO based connectivity for a 24 hour period.

Likewise, with the advent of 5G cellular technologies, the use of connectivity for critical vehicular functions will increase. Given that cellular does not and likely will not cover 100 percent of all roadways globally, The low likelihood that cellular service does not and likely will not cover 100 percent of all roadways globally underscores the need for a persistent and globally available back-up network. Our research demonstrates that a low earth orbit satellite network can help fill that gap in cellular connectivity. Early use cases might include scheduled software updates pushed to globally distributed vehicles, constantly available in-cabin entertainment, and the continuous uploading of vehicle telematics to the cloud.

The recent, worldwide COVID-19 pandemic only proves that the definition of the future of work must change. Our study demonstrates the readiness and feasibility of integrating low earth orbit (LEO) satellite as a provider of continuous connectivity in the mobility IT infrastructure. As the future of work is radically transformed, hybrid satellite-enabled infrastructure offers innovative pathways for resilient terrestrial communications.

## Acknowledgment

The authors would like to thank current and former staff at OneWeb, including Bala Srinivasan, Liang Shi, Frank Lopez, Jonathan Comstock, Rachana Kothare Bui-Pho, Rodrigo Gomez, Larry Alder, Jing Yan, and Heidi Dillard. The authors would also like to thank BMW staff, Ken Kennedy, Marcin Ziolowski, and Oliver Wick who were critical in pushing forward this work.